\documentclass[twocolumn,showpacs,preprintnumbers]{revtex4}
\usepackage{amssymb}
\usepackage{amsmath}
\usepackage{graphicx}
\usepackage{dcolumn}
\usepackage{bm}

\input{tcilatex}
\begin{document}

\title{The reversibility of the Goos-H\"{a}nchen shift near the
band-crossing structure of one-dimemsional photonic crystals containing
left-handed metamaterials}
\author{Li-Gang Wang$^{1,2}$ and Shi-Yao Zhu$^{1,2,3}$}
\affiliation{$^{1}$Centre of Optical Sciences and Department of Physics, The Chinese
University of Hong Kong, Shatin, N. T., Hong Kong\\
$^{2}$Department of Physics, Zhejiang University, Hangzhou, 310027, China\\
$^{3}$Department of Physics, Hong Kong Baptist University, Kowloon Tong,
Hong Kong}

\begin{abstract}
We perform a theoretical investigation on the Goos-H\"{a}chen (GH) shift in
one-dimensional photonic crystals (1DPCs) containing left-handed
metamaterials (LHMs). We find an unusal effect of the GH shift near the
photonic band-crossing structure, which is located at the condition, $%
-k_{z}^{(A)}d_{A}=k_{z}^{(B)}d_{B}=m\pi $ $(m=1,2,3\cdots )$, under the
inclined incident angle, here A denotes the LHM layer and B denotes the
dielectric layer. Above the frequency of the band-crossing point (BCP), the
GH shift changes from negative to positive as the incident angle increases,
while the GH shift changes reversely below the BCP frequency. This effect is
explained in terms of the phase property of the band-crossing structure.
\end{abstract}

\maketitle

It is well known that the Goos-H\"{a}chen (GH) shift refers to the lateral
shift of a well-collimated light beam totally reflected from the interface
of two different media \cite{Goos1947,Artmann1948}. Recently the GH shifts
have extensively studied in various situations, for example, dielectric slab
systems \cite{LiCF2002,LGWANG2005}, left-handed materials (LHMs) \cite%
{Berman2002,Lakhtakia2003}, Otto configuration \cite{Shadrivov2003b},
one-dimensional photonic crystals (1DPCs) \cite{Felbacq2003,LGWANG2006},
metal surfaces \cite{Leung2007}, 1DPCs with a nonlinear metamaterials \cite%
{Wei2008}, coherent driving atomic systems \cite{LGWANG2008}, and
electro-optic crystals \cite{ChenXi2008} and so on. Meanwhile, Li et al.
have discovered that multilayered structures containing LHMs possess zero
averaged refractive index gap (zero-$\overline{n}$ gap) \cite{LiJ2003},
which is quite different from a traditional Bragg gap. Shadrivov et al. \cite%
{Shadrivov2003} have further found that in such structures there exists
unusual angular dependencies of the beam transmission in which two photonic
bands may touch under a certain condition. In this report, we consider the
GH shifts near the band-crossing structure of the 1DPCs containing the LHMs,
and show that the GH shifts have unusual properties: Above the frequency of
the band-crossing point (BCP), the GH shift changes from negative to
positive as the incident angle increases, however the GH shift has the
opposite behavior below the BCP frequency.

\begin{figure}[tbp]
\centering
\includegraphics[width=7cm,trim=10mm 7.5mm 6mm 9mm,clip=true]{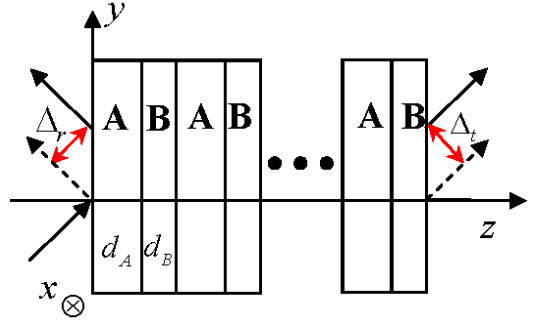}
\caption{Schematic of a $(AB)^{N}$ structures. $\Delta _{r}$ and $\Delta
_{t} $ are the reflected and transmitted GH shift.}
\label{fig:FIG1}
\end{figure}

For the simplicity, we only consider a TE-polarized light beam incident from
vacuum into the 1DPC at an incident angle $\theta $, as shown in Fig. 1. The
results are similar for TM-polarized light fields with frequencies near the
band-crossing structure. The 1DPC is consisted of alternative layers A and
B, with the structure $(AB)^{N}$, where $N$ is period. $d_{A}$ and $d_{B}$
are widths of layers $A$ and $B$, respectively. The layer A is a lossless
LHM with the effective dielectric permittivity and magnetic permeability 
\begin{equation}
\epsilon _{A}(\omega )=1-\frac{\omega _{ep}^{2}}{\omega ^{2}},\text{ }\mu
_{A}(\omega )=1-\frac{\omega _{mp}^{2}}{\omega ^{2}},  \label{EpsilonMu}
\end{equation}%
where $\omega _{ep}/2\pi =10$ GHz and $\omega _{mp}/2\pi =10.5$ GHz. The
layer B, without loss of generality, is vacuum ($\epsilon _{B}=1$ and $\mu
_{B}=1$) in this brief report. In general, the electric and magnetic fields
at any two positions $z$ and $z+\Delta z$ in the same layer can be related
via a transfer matrix \cite{WangLG2004}%
\begin{equation}
M_{j}(\Delta z,\omega ,\theta )=\left( 
\begin{array}{ll}
\cos \left[ k_{z}^{j}\Delta z\right] & i\frac{1}{q_{j}}\sin \left[
k_{z}^{j}\Delta z\right] \\ 
iq_{j}\sin \left[ k_{z}^{j}\Delta z\right] & \cos \left[ k_{z}^{j}\Delta z%
\right]%
\end{array}%
\right) ,  \label{Mmatrix}
\end{equation}%
where $k_{z}^{j}=\frac{\omega }{c}\sqrt{\varepsilon _{j}}\sqrt{\mu _{j}}%
\sqrt{1-\frac{\sin ^{2}\theta }{\varepsilon _{j}\mu _{j}}}$ is the $z$
component of wave vector $\vec{k}_{j}$ in the $j$th layer ($j=A,B$), $q_{j}=%
\frac{\sqrt{\varepsilon _{j}}}{\sqrt{\mu _{j}}}\sqrt{1-\frac{\sin ^{2}\theta 
}{\varepsilon _{j}\mu _{j}}}$, and $c$ is the light speed in vacuum. Then
the reflection and transmission coefficients [$r(\omega ,\theta )$ and $%
t(\omega ,\theta )$] can be readily obtained from the transfer matrix method 
\cite{WangLG2004}%
\begin{equation}
r(\omega ,\theta )=\frac{q_{0}(x_{22}-x_{11})-(q_{0}^{2}x_{12}-x_{21})}{%
q_{0}(x_{22}+x_{11})-(q_{0}^{2}x_{12}+x_{21})},  \label{rw1}
\end{equation}%
\begin{equation}
t(\omega ,\theta )=\frac{2q_{0}}{%
q_{0}(x_{22}+x_{11})-(q_{0}^{2}x_{12}+x_{21})},  \label{tw}
\end{equation}%
where $q_{0}=\cos \theta $ for the vacuum of the space $z<0$ before the
incident end and the space $z>L$ after the exit end ($L$ is the total length
of the 1DPC), and $x_{ij}$ ($i,j=1,2$) are the matrix elements of $%
X_{N}(\omega ,\theta )=\prod\limits_{j=1}^{2N}M_{j}(d_{j},\omega ,\theta )$
which represents the total transfer matrix for the finite 1DPC. For
obtaining the GH shift of the incident beam with a sufficiently large beam
waist (i. e., the beam with a very narrow angular spectrum, $\Delta k<<k$),
the GH shifts of both the reflected and transmitted beams can be expressed
explicitly as \cite{LGWANG2005}%
\begin{eqnarray}
\Delta _{r,t} &=&-\frac{\lambda }{2\pi }\frac{d\phi _{r,t}}{d\theta }  \notag
\\
&=&-\frac{\lambda }{2\pi }\frac{1}{|\Theta _{r,t}|^{2}}[\func{Re}(\Theta
_{r,t})\frac{d\func{Im}(\Theta _{r,t})}{d\theta }  \notag \\
&&-\func{Im}(\Theta _{r,t})\frac{d\func{Re}(\Theta _{r,t})}{d\theta }],
\label{GH Shift2}
\end{eqnarray}%
where $\phi _{r,t}$ are the phases of the reflection and transmission
coefficients, $\Theta _{r}=r(\omega ,\theta )$ and $\Theta _{t}=t(\omega
,\theta )$, and $\lambda $ correspond to the wavelength of the incident beam
with angular frequency $\omega $.

In order to know the information of the photonic band gap for an infinite
periodic structure ($N\rightarrow \infty $), according to Bloch's theorem,
the dispersion at any incident angle follows the relation \cite{Centini1999}%
\begin{eqnarray}
\cos \left[ \beta _{z}D\right] &=&\cos \left[
k_{z}^{(A)}d_{A}+k_{z}^{(B)}d_{B}\right] -\frac{1}{2}\left( \frac{q_{B}}{%
q_{A}}\right.  \notag \\
&&\left. +\frac{q_{A}}{q_{B}}-2\right) \sin \left[ k_{z}^{(A)}d_{A}\right]
\sin \left[ k_{z}^{(B)}d_{B}\right] ,  \label{BlochVector}
\end{eqnarray}%
where \bigskip $\beta _{z}\ $is the $z$ component of Bloch wave vector, and $%
D=d_{A}+d_{B}$. The condition of Eq. (\ref{BlochVector}) having no real
solution for $\beta _{z}$ is $\left\vert \cos [\beta _{z}D]\right\vert >1$,
which is well-known as the Bragg condition of the photonic band gap. In the
LHMs (i. e., layers A) $k_{z}^{(A)}$ is negative, while in layers B $%
k_{z}^{(B)}$ is positive. Therefore Eq. (\ref{BlochVector}) becomes%
\begin{eqnarray}
\cos \left[ \beta _{z}D\right] &=&\cos \left[
k_{z}^{(B)}d_{B}-|k_{z}^{(A)}d_{A}|\right] +\frac{1}{2}\left( \frac{q_{B}}{%
q_{A}}\right.  \notag \\[0.15cm]
&&\left. +\frac{q_{A}}{q_{B}}-2\right) \sin \left[ |k_{z}^{(A)}d_{A}|\right]
\sin \left[ k_{z}^{(B)}d_{B}\right] .  \label{BlochVector2}
\end{eqnarray}%
When $-k_{z}^{(A)}d_{A}=k_{z}^{(B)}d_{B}\neq m\pi $ ($m=1,2,3,\cdots $), Eq.
(\ref{BlochVector2}) has no real solution. Therefore there exists a
distinctive band gap, the so-called zero-averaged refractive index gap (zero-%
$\overline{n}$ gap) \cite{LiJ2003}. However, when%
\begin{equation}
-k_{z}^{(A)}d_{A}=k_{z}^{(B)}d_{B}=m\pi \text{ \ }(m=1,2,3,\cdots ),
\label{condition}
\end{equation}%
the upper and lower photonic bands may touch together. This phenomena has
been pointed by Shadrivov et al. in Ref. \cite{Shadrivov2003}. Here we would
like to emphasize that, in the case of normal incidence, Eq. (\ref{condition}%
) leads to the touch effect of both the upper and lower photonic bands,
however such an effect does not belong to the band-crossing effect because
the upper and lower bands are still parabolic shape. In the case of inclined
incidence, Eq. (\ref{condition}) leads to the band-crossing effect, which
refers to that both the upper and lower photonic bands touch together and
the dispersion near the touch point is linear. In the following discussion,
we focus ourselves to investigate the GH shift near the band-crossing
structure when the 1DPC's structure satisfies the condition of Eq. (\ref%
{condition}) in the case of inclined incidence.

\begin{figure}[tbp]
\centering
\includegraphics[width=8cm,trim=75mm 2mm 0mm 5mm,clip=true]{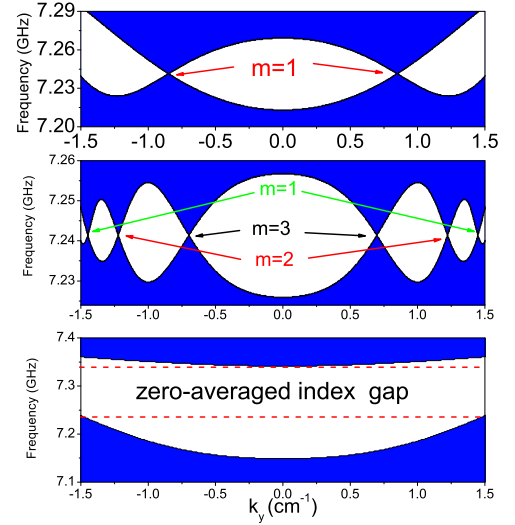}
\caption{(Color online). Photonic band structures of infinite $(AB)^{N}$
structures: (a) $d_{A}=d_{B}=2.5$cm, (b) $d_{A}=d_{B}=7$cm, and (c) $%
d_{A}=d_{B}=1.2$cm. Heavy solid areas correspond to the pass bands, and
white areas are the photonic band gaps.}
\label{fig:FIG2}
\end{figure}

In Fig. 2, we plot the photonic band structure in the parameter plane ($%
\omega ,k_{y}$), where the transverse wave number $k_{y}=\frac{\omega }{c}%
\sin \theta $ is relative to incident angle $\theta $ for a fixed $\omega $.
From Fig. 2(a) and (b), it is seen that the band-crossing effect occurs at
the angles where Eq. (\ref{condition}) is satisfied for the inclined
incidences. Close to the BCP, the dispersion relation between $\omega $ and (%
$k_{y}-k_{ym}$) is linear, where $k_{ym}$ is the transverse wave number for
the $m$th band-crossing structure. We find that such band-crossing
structures are similar to the Dirac band structure in some two-dimensional
photonic crystals, which is consisted of triangular or honeycomb lattices of
the dielectric cylinders \cite%
{Plihal1991,add2007,Haldane2008,Sepkhanov2007,Zhang2008}. Recently, we also
found that it is possible to realize such a band-crossing structure in
homogenous negative-zero-positive index media \cite{LGWang2009a}. Near the
band-crossing structure, one has demonstrated some unusual properties, such
as conical diffraction \cite{add2007}, static diffusive property of the
light fields \cite{Sepkhanov2007,LGWang2009a} and \textit{Zitterbewegung} of
optical pulses \cite{Zhang2008,LGWang2009b}. For comparison, when $%
-k_{z}^{(A)}d_{A}=k_{z}^{(B)}d_{B}<\pi $, we plot Fig. 2(c) to demonstrate a
zero-$\overline{n}$ gap, which is an omnidirectional gap and is almost
independence of incident angle \cite{HTJiang2003}.

\begin{figure}[tbp]
\centering
\includegraphics[width=8cm,trim=0mm 2mm 0mm 35mm,clip=true]{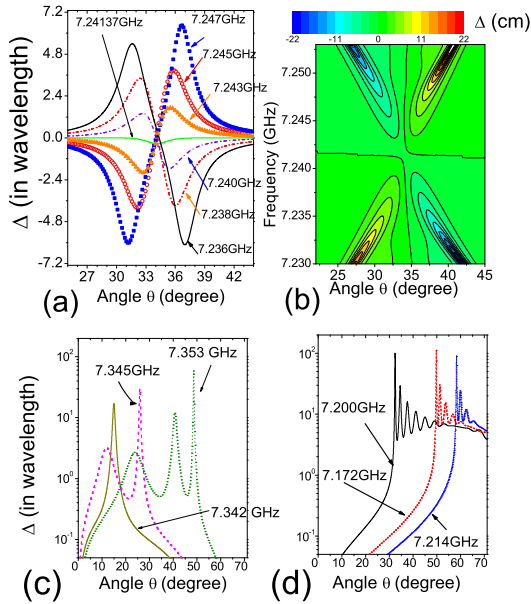}
\caption{(Color online). (a) The typical GH shift $\Delta $ as a function of
angle $\protect\theta $ near the band-crossing structure with $m=1$. (b)
Contour plot of (a) as functions of frequency and angle. The GH shifts near
the band edges of (c) the lower passing band and (d) the upper passing band
at both sides of the zero-$\overline{n}$ gap in Fig. 2(c).}
\label{fig:FIG3}
\end{figure}

In Fig. 3(a), we plot the typical GH shifts of transmitted light beams near
a band-crossing structure with $m=1$, as shown in Fig. 2(a). The results for
the reflected light beams are similar and are not plotted in all following
cases. Fig. 3(a) shows that, above the BCP frequency $(\omega >\omega
_{BCP}, $ in our cases $\omega _{BCP}\approx 2\pi \times 7.24137$ GHz$)$,
there is a negative GH shift as incident angle $\theta $ transits from the
gap region into the passing band region, however it becomes a positive GH
shift as $\theta $ increases from the passing band into the gap region.
Therefore the GH shift for the light with frequencies above the BCP
frequency can change from negative to positive with the increasing of $%
\theta $. Conversely, below the BCP frequency $(\omega <\omega _{BCP})$, the
GH shift changes from positive to negative with the increasing of $\theta $%
.\ Figure 3(b) is a contour plot of Fig. 3(a). It is clear that, near the
band-crossing structure, the GH shift can be changed from positive to
negative or from negative to positive, depending on light frequency located
below or above the BCP frequency. For comparison, we plot Figs. 3(c) and
3(d) to show the GH shifts in the case of Fig. 2(c). It is seen that the GH
shifts near the band edges of the zero-$\overline{n}$ gap are always
positive, which is similar to that in the conventional 1DPCs \cite%
{Felbacq2003} but is very different from the present cases. Therefore we
conclude that the GH shifts inside the band-crossing structures are very
unusual: it can change from negative to positive above the BCP frequency and
change conversely below that frequency. We can expect that there are similar
GH effects near the band-crossing structures with $m=3$ and $m=2$, see Fig.
4(a) and 4(b), which correspond to the cases in Fig. 2(b). From Fig. 3(a),
and Figs. 4(a) and 4(b), it is further found that the GH shift changes
almost linearly from positive to negative (or from negative to positive)
inside the passing bands of such band-crossing structures.

\begin{figure}[tbp]
\centering
\includegraphics[width=9cm]{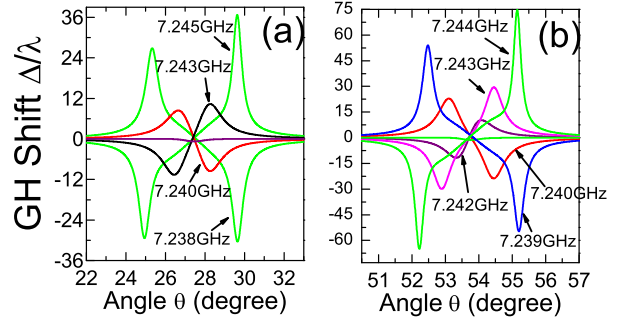}
\caption{(Color online). The GH shift $\Delta $ near the band-crossing
structures with (a) $m=3$ and (b) $m=2$ as in Fig. 2(b).}
\label{fig:FIG4}
\end{figure}

\begin{figure}[bp]
\centering
\includegraphics[width=8cm,trim=0mm 2mm 0mm 5mm,clip=true]{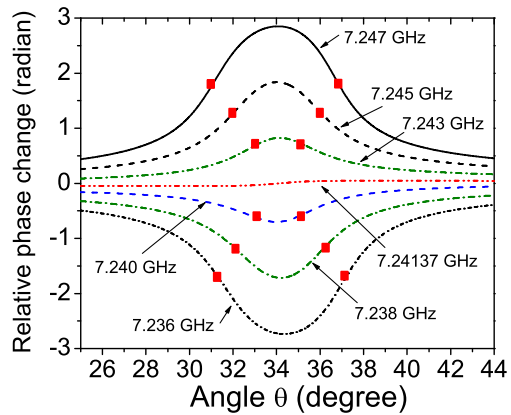}
\caption{(Color online). The relative phase shift as a function of angle $%
\protect\theta $ at both sides of the band-crossing structure with $m=1$. On
each curve, the range between two solid squared points denotes the range of
the passing band.}
\label{fig:FIG5}
\end{figure}

To gain a deeper insight into the physical mechanism for such reversed GH
shifts near the band-crossing structures, we plot Fig. 5 to show the
relative phase change as a function of $\theta $ at different frequencies.
From Fig. 5, the relative phase change of the transmitted light beam,
respect to the phase of the incident light beam, varies from negative to
positive as the light frequency crosses over the BCP frequency. Here we have
to point out that the BCP frequency corresponds to the frequency with a
zero-averaged refractive index of the total 1DPC. The relative phase is
always positive above the BCP frequency; meanwhile it varies initially from
a small to large positive value when $\theta $ moves from the gap region
into the passing band and then decreases from the maximal phase change to a
small one when $\theta $ moves out of the passing band again. Thus the GH
shifts for the light beams with frequencies above the BCP frequency change
from negative to positive, and the GH shifts can be very large negative (or
large positive) near the transition regions from the gap region (or the
passing band) to the passing band (or the gap region). From Fig. 5, it is
clear that below the BCP frequency, the relative phase is completely
opposite with that above the BCP frequency. Therefore the GH shift is
reversed near the band-crossing structure when the light frequency is
located below the BCP frequency. Finally, it should be pointed out that the
phase change with respect to $\theta $ is a quadratic-like curve inside the
passing band (see the regions between two squared points on each curve in
Fig. 5), therefore the shifts inside the passing bands change linearly from
negative to positive above the BCP frequency (or from positive to negative
below the BCP frequency).

In summary, we have investigated the GH shifts in 1DPCs containing LHMs. It
is found that there is an unusual effect for the GH shift near the photonic
band-crossing structure, which appears at the resonant condition $%
-k_{z}^{(A)}d_{A}=k_{z}^{(B)}d_{B}=m\pi $ $(m=1,2,3\cdots )$ at the inclined
incident angle. Above the BCP frequency, the GH shift changes from negative
to positive as $\theta $ increases, while the GH shift changes reversely
below the BCP frequency. Such unusual GH shifts are helpful for
understanding on the special properties of the band-crossing structures.
Finally we have explained this effect in terms of the phase properties of
the upper and lower passing band of the band-crossing structure.

\textbf{Acknowledgments}: This work is supported by NSFC05-06/01, CUHK
401806 and 2060360, and FRG of HKBU, and NSFC (10604047).

\end{document}